\documentclass[fleqn,twoside]{article}
\usepackage{espcrc2}

\usepackage{graphicx}
\usepackage{epsfig}
\usepackage[figuresright]{rotating}

  \def\dm{\Delta m^2}
 
\newcommand{\AmS}{{\protect\the\textfont2
  A\kern-.1667em\lower.5ex\hbox{M}\kern-.125emS}}

\title{Future projects on atmospheric neutrinos}

\author{T. Tabarelli de Fatis\address[INFN]{Istituto Nazionale di
        Fisica Nucleare, Sezione di Milano \\
        Piazza della Scienza 3, I-20126 Milano, Italy}}
       
\begin{document}

\begin{abstract}
New results from Super-Kamiokande, K2K and SNO not only have spurred on
the interest in neutrino oscillation physics, but also have started to
shift the interest from discovery to precision measurements. Future
projects focusing on atmospheric neutrinos are reviewed in this
context. Important contributions could be made in the precision
determination of the oscillation parameters, in the observation of
matter effects and in the determination of the neutrino mass hierarchy. 
Unfortunately, the probability that the projects discussed in this
review will be running in the next ten years is rather small. The
only project with a shorter time scale has not been funded. 

\vspace{1pc}
\end{abstract}

\maketitle

\section{Introduction}

The growing evidence of oscillation in atmospheric \cite{SK,SK-2002} 
and solar \cite{SK-sol,SNO} neutrino experiments, corroborated by the 
first long base-line accelerator experiment \cite{K2K}, is shifting the 
interest 
in neutrino physics
from discovery to stoichiometry, i.e. precision measurements
of neutrino masses and mixings.

A pure $\nu_\mu \to \nu_\tau$ oscillation economically describes 
all current observations of atmospheric neutrinos. In this context,
future projects on atmospheric neutrinos could significantly improve 
the precision on the oscillation parameters through the explicit
observation of the as-yet elusive oscillation pattern in the $L/E$
spectrum\footnote{$L$ and $E$ are neutrino baseline and its energy} of
atmospheric muon neutrinos. This would also provide a direct proof of
oscillations.  

The three neutrino scenario, needed to fit simultaneously atmospheric 
and solar neutrino data, is even less constrained. In particular, 
a sub-dominant contribution of $\nu_e$ to atmospheric neutrino
oscillations, related to a non vanishing value of the  mixing angle
$\theta_{13}$, is not excluded by data \cite{SK-2002,CHOOZ}. The 
$\nu_\mu \to \nu_e$ transition probability could be sizeable amplified  
in particular regions of the spectrum through earth-induced matter 
effects. Depending on the sign of $\Delta m^2$, these 
effects occur either for neutrinos or for anti-neutrinos only. 
The study of earth-induced matter effects is thus a topic of major 
interest for future projects on atmospheric neutrinos, since it
provides an handle to measure or constrain $\theta_{13}$ and 
to determine the neutrino mass hierarchy. 

More complex scenarios, e.g. with additional (sterile) neutrino, 
would be needed to accommodate also the LSND results \cite{LNSD}. 
These will not be covered by the present review. Still, one should be
open to unexpected results in future and  more precise experiments.

\section{The atmospheric neutrino beam}

Atmospheric neutrinos are 
characterized by a wide $L/E$ spectrum (from about 1 km/GeV to 10$^5$
km/GeV). This not only give access to a very large range of oscillation
parameters down to small $\dm$, but is also crucial to constrain non
$L/E$ contributions to the transition probability \cite{LipLus}, 
expected from sub-dominant effects, like matter effects or non standard 
interactions. 

Moreover, the very long base-lines available with atmospheric
neutrinos offer the possibility to search for earth-induced matter
effects. In that  endeavour, atmospheric neutrino experiments are not
contested by current accelerator beam programs, whose 
base-lines (250$\div$730~km) are too short for a significant effect. 

Above about 1~GeV of neutrino energy, the flux of atmospheric
neutrinos is up/down symmetric at the 1\%
level \cite{Gai98}. This is an ideal condition for disappearance 
experiments and gives the opportunity to perform oscillation studies
with little systematic uncertainties from the knowledge of the beam
spectrum: down-going neutrinos constitute a {\it near} reference
source to which compare the {\it far} source of up-going neutrinos. 

The atmospheric neutrino beam has an approximately equal content of
neutrinos and anti-neutrinos, which is useful for matter effect
studies, if the events can be charge-classified. On the other hand,
although the ratio of the $\nu_e$ to $\nu_\mu$ flavours is known with 
reasonable approximation  \cite{Gai98}, the almost democratic flavour
composition of the beam is not ideal for appearance experiments.  

Due to the low intensity of the atmospheric neutrino flux, the main
limitation of atmospheric neutrino experiments stems from the limited
statistics, in particular at high energies. Decisive progresses
require massive detectors. 

\section{Future atmospheric neutrino detectors}

The reference against which future progress should be evaluated is
given by the (not yet final) results of Super-Kamiokande, a 50 kt 
(23 kt fiducial) water Cherenkov detector, that is expected to run 
still for several years.
 
\subsection{Water Cherenkov detectors}

Some of the atmospheric neutrino measurements are limited by the
available statistics, in particular of high energy neutrino events,
and could be improved by extending the Super-Kamiokande concept to a
larger detector mass. 

\begin{figure}
\begin{center} 
\epsfig{file=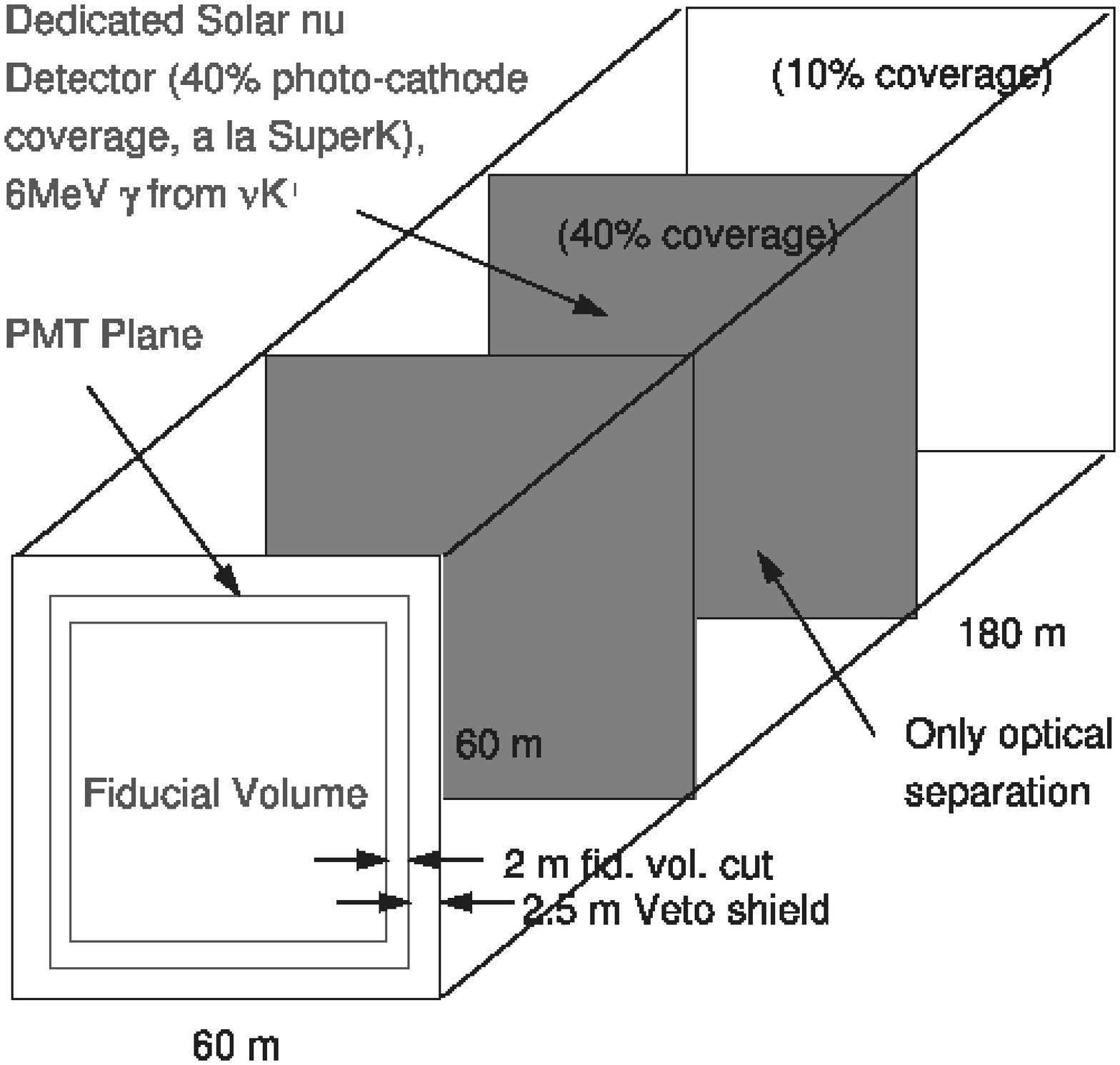, width=\linewidth}
\caption{The UNO detector concept.}
\label{figUNO}
\end{center}
\end{figure}

A 650 kt (450 kt fiducial) Water Cherenkov detector, tentatively 
called UNO  
has been proposed \cite{UNO}. In addition to atmospheric
neutrinos, this detector is intended to also address the search of 
nucleon decay, the detection of solar and supernova neutrinos and the
detection of neutrinos from long base-line beams. 
The UNO detector would consist of three cubic compartments of
$60 \times 50 \times  60$~m$^3$ (Fig.\ref{figUNO}). The central
compartment would be equipped with photo-multiplier with the same
coverage as in Super-Kamiokande, to be fully sensitive also to low
energy events (in the 10 MeV range). A less dense photo-multiplier
coverage is foreseen for the two outer compartments, dedicated to high
energy events.  

A similar project (Hyper-Kamiokande) is discussed in
Japan as candidate successor of Super-Kamiokande. It would be a 1~Mt 
(800~kt fiducial) water Cherenkov detector, made of eight cubic
compartments of $50 \times 50 \times50$~m$^3$, aligned along the JHF2K
beam direction. The photo-multiplier coverage and design are
still under study \cite{JHF}.

Water Cherenkov detectors rely on an experimental technique
extensively tested by Super-Kamiokande and, to a large extent, on
known technologies. Moreover, the detector is relatively cheap (around 
0.5 MEuro/kt, with an additional 0.5 MEuro/kt for excavation costs). 
Some drawbacks are the time scale for their realization, which spans
at list a decade, and the difficulty to implement a magnetic field to
measure the muon charge and to extend the acceptance at high energies 
with semi-contained events.

On a longer times scale, a different detector concept was
envisaged by the Aqua-RICH project \cite{Anton}. In this approach,
the use of the ring imaging technique would provide a cleaner particle  
identification and the measurement of the particle momentum for
outgoing tracks from multiple scattering. This detector, however,
still require a major R\&D phase. 


\subsection{Magnetised Iron Neutrino Detectors}

A different detector concept is represented by large magnetised iron
calorimeters, which offer the advantage of the measure of the muon
charge: a very crucial feature to test matter effects with atmospheric 
neutrinos. 

Moreover, for masses comparable to Super-Kamiokande, these detectors 
would have a much larger acceptance to high energy muons, where the 
muon direction gives a better estimate of the neutrino direction. 
This results in a considerably improved $L/E$ resolution and overcomes
the main limitation in the sensitivity to the oscillation pattern of
current atmospheric neutrino experiments. 

One such detector, the approved MINOS detector at Fermilab
\cite{Minos}, has been primarily designed as long base-line
beam detector. Its limited fiducial mass (about 3.5 kt) severely
limits its potential contribution to atmospheric neutrino
measurements. 

\begin{figure}
\begin{center} 
\epsfig{file=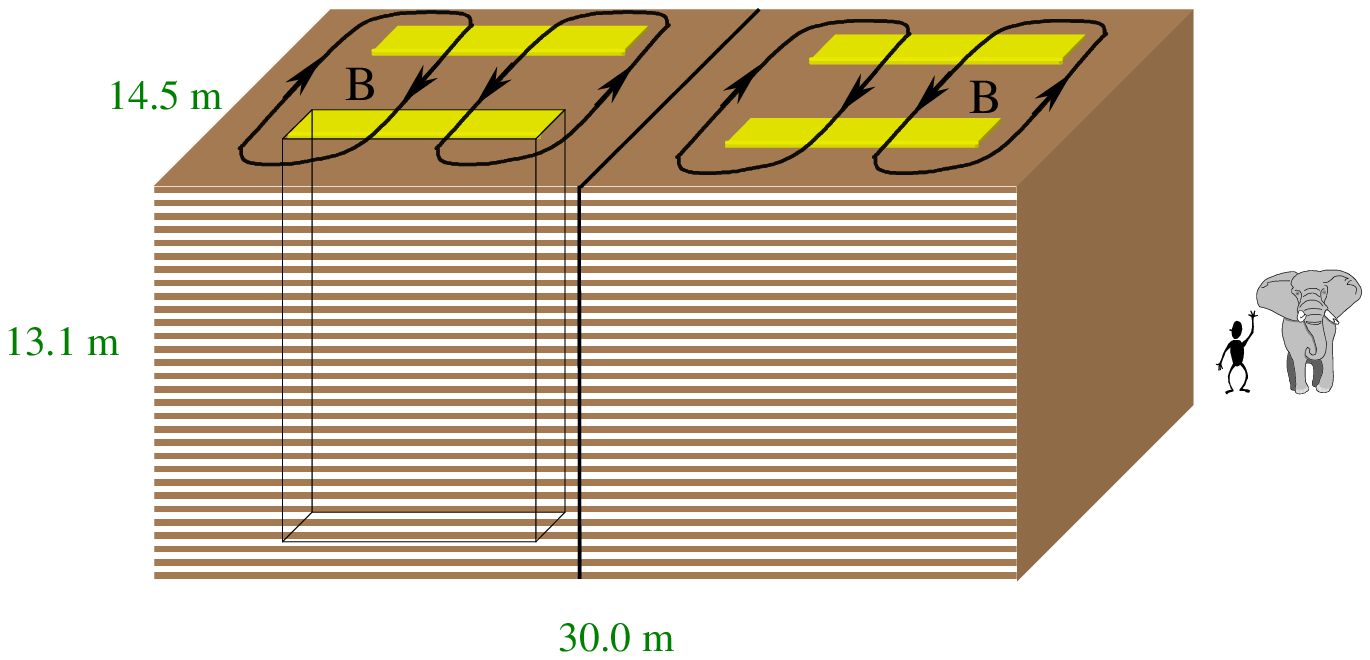, angle=-90, width=\linewidth}
\caption{Schematic view of the MONOLITH detector. The arrangement 
         of the magnetic field is  also shown.} 
\label{fig:module}
\end{center}
\end{figure}

A different concept was proposed for MONOLITH \cite{Monolith}, a
massive tracking calorimeter with coarse structure and intense
magnetic field (about 1.3~T), explicitly designed for oscillation
studies with atmospheric muon neutrinos in the Gran Sasso laboratory
in Italy. The detector has a large modular structure 
(Fig.\ref{fig:module}). One module consists in a stack of 125
horizontal 8~cm thick iron planes with a surface area of $14.5\times
15\ {\rm m^2}$, interleaved with 2 cm planes of sensitive
elements, which provide a two-coordinate readout with good time
resolution. The total mass of the detector for two modules is about
34~kt. The unit cost of this detector (about 1 MEuro/kt) is comparable 
to that of large water Cherenkov detectors, excavation 
costs are included. 

The MONOLITH project was closed by INFN in fall 2001. However, a
similar detector is being considered by the INO collaboration. 
This detector would be located in India, where it could also possibly
act as a far-end detector of a long base-line beam experiment from
Japan \cite{Gandhi}.

Magnetised iron neutrino detectors with masses of order 100~kt 
are also candidate detectors for a future neutrino
factory \cite{nufcalo}. Their performance on atmospheric neutrinos
would be similar to MONOLITH but on a longer time scale. 

\subsection{Liquid Argon TPC's}

Liquid Argon Time Projection Chambers (LAr TPC's) are also well suited to 
the detection of atmospheric neutrinos. The interest of these
detectors stems from their superior resolution and their possibility
to flavour-classify the events. With the present technology, their
cost is much higher (about 10 MEuro/kt) than for iron calorimeters or
water Cherenkov detectors. The implementation of a magnetic field is
also not obvious.

A 600 t LAr TPC detector is being installed by the ICARUS
collaboration in the Gran Sasso laboratory \cite{ICARUS}. A proposal
to increase the mass up to 3.0~kt with additional modules has been 
submitted \cite{ICAR2}. The main focus of this proposal is on the
detection of the $\nu_\tau$ and $\nu_e$ appearance in the CNGS beam
from CERN to Gran Sasso \cite{CNGS} and it is reviewed in other
contributions to this conference \cite{Katsa}. Its potential
contribution to atmospheric neutrino measurements is limited by the
small mass. However, for some measurements, its superior  resolution
could compensate the reduced statistics.  

In the long term, large Liquid Argon TPC's with masses in excess of 30
kt have also been considered at the Letter of Intent level
\cite{SuperIcarus,LAND}. 

\section{Precision measurement of atmospheric $\nu_\mu$ disappearance}

\begin{figure}
\begin{center}
\mbox{\epsfig{file=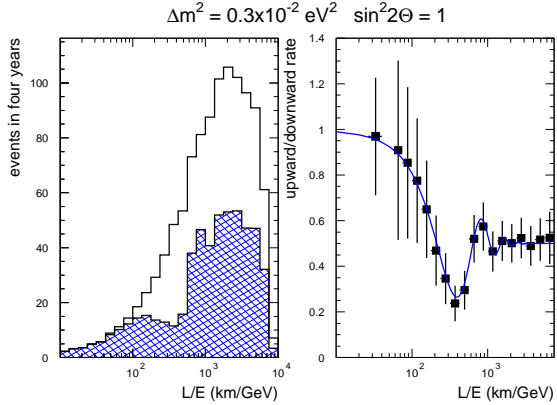,width=\linewidth}} 
\end{center}
\caption{
  Left: The $L/E$ spectrum of up-going (hatched) and the reference
  spectrum of down-going (open histogram) muon neutrino events. Right:
  Their ratio with the best-fit of $\nu_\mu$--$\nu_\tau$ oscillation
  superimposed. Rates and errors are normalized to 4 y of MONOLITH 
  exposure \cite{Monolith,ttf01}. 
} 
\label{fig:one}
\end{figure}

The up/down asymmetry of high energy atmospheric muon neutrinos
observed by Super-Kamiokande, gives a 10$\sigma$ evidence for neutrino
oscillations. This also result in a 10\% precise determination of
$\sin^2 2\theta_{23}$, fully dominated by the statistical error
\cite{SK-2002}. On the other hand, due to the limited $L/E$ resolution, 
Super-Kamiokande does not have so far succeeded to measure an
oscillation pattern in the $L/E$ spectrum of the atmospheric muon
neutrinos. Thus, a direct proof of oscillations is still outstanding
and, most importantly, the measurement of $\dm$ is not accurate.

As anticipated, the much larger acceptance to high energy muons of 
massive magnetised iron detector, hence their improved $L/E$
resolution, is particularly rewarding. MONOLITH, with a mass comparable
to Super-Kamiokande and after a similar exposure, could already
resolve the full first oscillation swing in the $L/E$ spectrum. 
The expected $L/E$ distributions of up-going neutrinos and of the
reference sample of down-going neutrinos, which are assigned the
baseline $L$ they would have travelled if they were produced with a
nadir angle equal to the observed zenith angle \cite{Pic97}, are
shown in Fig.\ref{fig:one} for $\nu_\mu$--$\nu_\tau$ oscillations,
with $\Delta m^2 = 3\times 10^{-3}$ eV$^2$ and $\sin^2
2\theta=1.0$. In the ratio of the two distributions the oscillation
behaviour is evident.  

\begin{figure}
\begin{center}
\mbox{\epsfig{file=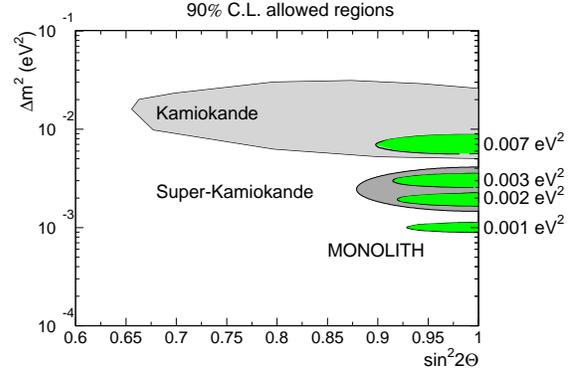,width=\linewidth}}
\end{center}
\caption{  
  Expected allowed regions at 90 \% C.L. for $\nu_\mu$--$\nu_\tau$
  oscillations after 4 y exposure. 
  The results for $\Delta m^2 = 1, 2, 3, 7 \times 10^{-3}$ eV$^2$
  and maximal mixing are compared to the Super-Kamiokande 
  and Kamiokande allowed region \cite{ttf01}.
}
\label{sensibilita}
\end{figure}

The clearness of the oscillation pattern results in a significantly
improved measurement of the oscillation parameters
(Fig.\ref{sensibilita}), with a precision on $\Delta m^2$ and $\sin^2
2\theta$ around 6\%. The observation of a first clear dip in the
$L/E$ distribution would also enable MONOLITH to disprove
unconventional interpretations \cite{decay,decoherence} of
Super-Kamiokande data. 

The sensitivity range for this measurement comfortably covers the
entire region of parameters allowed by current results and does not
strongly depend on the oscillation parameters. This is in
contrast to long base-line experiments like MINOS and K2K, for which 
the observation of a full oscillation swing, including $\nu_\mu$
``reappearance'', in the low $\dm$ range is not obvious. 

At this level, the precision of this flux independent method is
limited only by statistics. The ultimate sensitivity on the mixing
angle is about 1\%, due to deviations from the exact up/down
symmetry of the atmospheric neutrino flux at high energy. 
An uncertainty of a few per cent on $\dm$ is expected from 
the uncertainty on the calibration of the $L/E$ scale and on the
resolution function.  

\begin{figure}
\begin{center}
\mbox{\epsfig{file=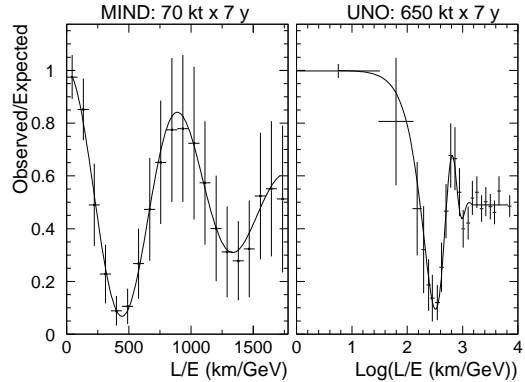,width=\linewidth}}
\end{center}
\caption{  
    Expected oscillation pattern in MIND (left, adapted from  
    Ref. \cite{Geiser00}) and in UNO (right, adapted from
    Ref. \cite{UNO}).  
    Both graphs are normalised to 7 y of exposure.
}
\label{minduno}
\end{figure}

Two examples of very large exposure are shown in Fig.\ref{minduno}, 
where the $L/E$ pattern expected in a Magnetised Iron Neutrino
Detector (MIND) of about twice the MONOLITH mass (70~kt) and in the 
UNO detector (650~kt) are compared. In these detectors, higher
resolution events samples could be selected, yielding an improved 
clearness of the $L/E$ pattern. Due to its lower acceptance at high
energies, the UNO detector needs an exposure about ten times larger
then MIND to achieve the same accuracy in the reconstruction of the
oscillation pattern. 
In other words, the plot on the right-hand side of the figure is about
ten times more expensive than the plot on the left-hand side. 

\section{Search for earth-induced matter effects}

\subsection{Matter effects from $\nu_e \to \nu_\mu$ transition at
the atmospheric $\dm$}

\begin{figure}[t]
\begin{center}
\mbox{\epsfig{file=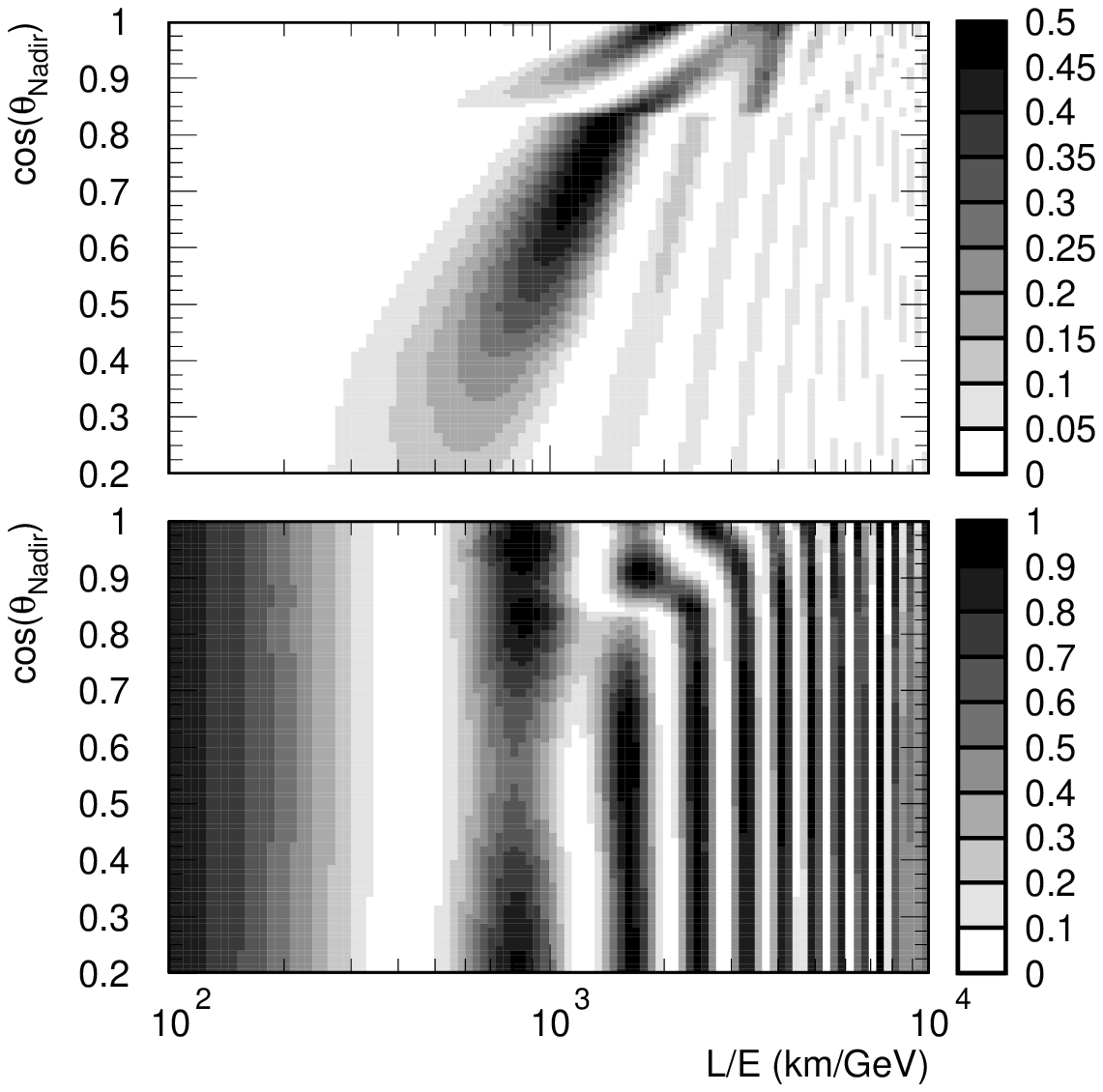,width=0.99\linewidth}}
\end{center}
\caption{$P(\nu_e \to \nu_\mu)$ (top) and $P(\nu_\mu \to \nu_\mu)$
(bottom) for neutrinos crossing the earth as a function of $L/E$ and
$\cos\theta_{Nadir}$. The calculation is presented for 
$\dm = +0.003$~eV$^2$ and $\sin^2 2\theta_{13}=0.1$. Anti-neutrinos, 
not shown in the figure, are not affected by matter effects for
positive $\dm$.} 
\label{probosc}
\end{figure}

In standard three flavour scenarios, matter effects are present only
for non vanishing mixing angle $\theta_{13}$. This parameter is bound
to be small by CHOOZ results \cite{CHOOZ}, 
but the $\nu_e \to \nu_\mu$ transition 
can become resonant in matter and sizeably modify the oscillation
probabilities of electron and muon neutrinos. An example is given in
Fig.\ref{probosc}, for $\dm = +0.003$~eV$^2$ and $\sin^2
2\Theta_{13}=0.1$, the largest valued allowed at 90\% C.L. by CHOOZ. 
The distortions due to matter resonant
transitions are particularly evident in the earth's mantle, for
energies around 7 GeV and a base-line of about 9000~km. Depending on
the sign of $\Delta m^2$, these effects occur either for neutrinos or
for anti-neutrinos only. The sign of $\Delta m^2$, i.e. the neutrino
mass hierarchy, could be determined, if  the contributions of
neutrinos and anti-neutrinos could be separated. The size of this
effect would measure the admixture of electron neutrinos.  

Since both electron an muon neutrinos are present in the atmospheric 
neutrino flux the transition probabilities shown in Fig.\ref{probosc} 
are not directly accessible to experiments. The observable quantity 
is the ratio of the observed to the expected rates of neutrino 
events of a given flavour,  which is a combination of the conversion 
probabilities displayed in Fig.\ref{probosc}, weighted by the
corresponding neutrino fluxes. The observable effects are less
impressive than those shown in the figure.

\begin{figure}
\begin{center}
\mbox{\epsfig{file=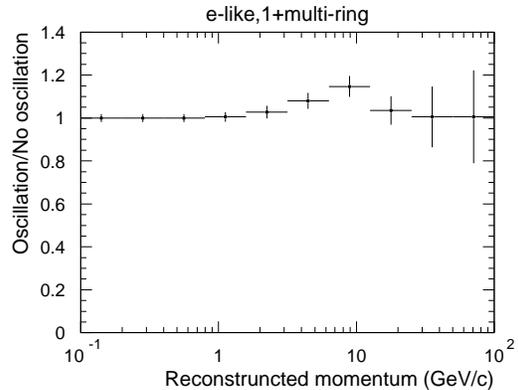,width=\linewidth}}
\end{center}
\caption{Ratio of up-going {\it e-like} events in case of three
neutrino oscillation with $\sin^2 2 \theta_{13}=0.1$ to the
expectations for pure $\nu_\mu \to \nu_\tau$ oscillations. 
Errors are normalized to an exposure of 900 kty (adapted from
Ref. \cite{Kajita01}).} 
\label{kaji}
\end{figure}

In $\nu_e$ appearance measurements \cite{Kajita01}, 
electron neutrino events can hardly be charge-classified and, 
in a water Cherenkov detector, only the inclusive spectrum of {\it e-like} 
events can be measured. The expectations for about 1~y of
Hyper-Kamiokande exposure are shown in Fig.\ref{kaji}, assuming
$\sin^22\theta_{13}$ at the  CHOOZ limit and positive sign of $\dm$.  
The resonance height would be about two times lower for negative $\dm$, 
since only about 1/3 of the {\it e-like} sample is originated by
anti-neutrino interactions.   
The height of the resonance is also affected by the value of the mixing
angle and the two quantities cannot be disentangled. If the
mixing angle is fixed by some future long base-line experiment, the
resonance height could bring information on the sign of $\dm$.
With a exposure of 900~kty, the separation between $\pm \dm$
hypotheses is about at the 2$\sigma$ level, for the maximum allowed
value of $\theta_{13}$. 


\begin{figure}[t]
\begin{center}
\mbox{\epsfig{file=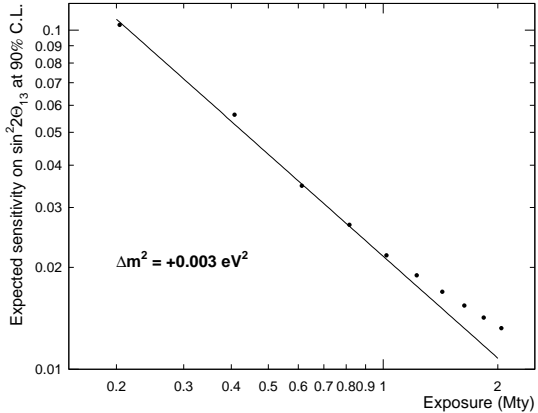, width=\linewidth}}
\end{center}
\caption{ Expected upper limits at 90\% C.L. as a function of the
exposure, if  no matter  effects are observed  in MIND. $\dm =
+0.003$ eV$^2$ is assumed. The line shows the  scaling
behaviour expected from statistical fluctuations only \cite{ttf02}. 
} 
\label{long}
\end{figure}

More promising, if the value of $\sin^2 2\theta_{13}$ is not too
small, seem the prospects of measuring $\sin^2
2\theta_{13}$ and the sign of $\Delta m^2$ with a magnetised iron
detector, where the $\nu_\mu$ disappearance can be studied separately
for neutrinos and anti-neutrinos \cite{ttf02}. 
The systematic uncertainties related to the knowledge of the
atmospheric neutrino rates may be controlled using the reference
source of down-going neutrinos, while the electron density profile in
the earth's mantle, where matter effects are observable, is known with 
sufficient precision.   
The sensitivity to  $\sin^2 2\theta_{13}$ is thus limited only by the
available statistics up to an exposure of about 1~Mty, corresponding
to a sensitivity around 0.02 on the mixing parameter (Fig.\ref{long}). 
Beyond that limit, resonant effects would produce spectral distortions
comparable to the experimental resolution and cannot be resolved. 

\begin{figure}[t]
\begin{center}
\mbox{\epsfig{file=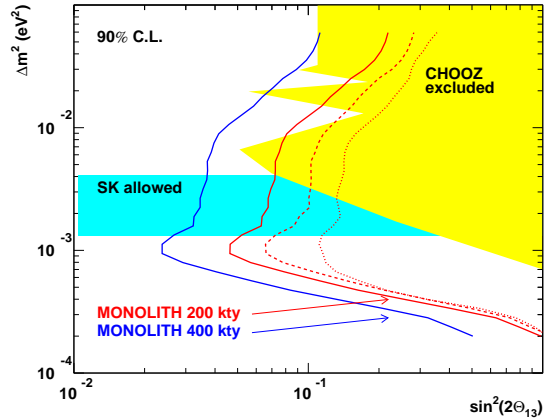,width=0.98\linewidth}}
\end{center}
\caption{
  Regions of oscillation parameters over which the sign of $\Delta
  m^2$ can be determined at the 90\% C.L. after MONOLITH exposures of
  200~kty (6 y),  assuming that $\sin^2 2\theta_{13}$ be known with
  30\% accuracy (full line), that $\sin^2 2\theta_{13}$ be unknown and  
  $\dm$ positive (dashed) and  that $\sin^2 2\theta_{13}$ be unknown
  and  $\dm$ negative. The first curve is also reproduced for
  400~kty. The regions excluded by CHOOZ and allowed by
  Super-Kamiokande at 90\% C.L. data are also shown \cite{ttf02}. 
} 
\label{matm}
\end{figure}

The range of sensitivity to $\sin^2 2\theta_{13}$ achievable with
atmospheric neutrinos will be most probably covered by projects
searching for the direct $\nu_\mu \to \nu_e$ transition with
conventional neutrino beams, like MINOS \cite{Para}, ICARUS
\cite{ICAR2} and JHF2K \cite{JHF}. 
Even with positive evidence, however, 
these experiments can not measure the sign of $\Delta m^2$,
which can be tested with atmospheric neutrinos by comparing neutrino
to anti-neutrino spectra. The sensitivity to the sign of $\Delta m^2$
is shown in Fig.\ref{matm} for MONOLITH and under different
assumptions on a potential prior knowledge of $\sin^22\theta_{13}$. 
For unknown $\sin^22\theta_{13}$, the expected sensitivity is about 
two times better for positive than for negative sign of $\dm$, due to
the larger event rate of atmospheric neutrinos than anti-neutrinos.
%
%
%
Assuming that $\sin^22\theta_{13}$ be determined with
30\% precision at future accelerator experiments, an
exposure of 200 kty (6~y) would be sufficient to fully evade the CHOOZ
exclusion region. The sensitivity scales according to Fig.\ref{long}, 
with an ultimate sensitivity for $\sin^2 2\theta_{13} \simeq 0.01$ at very large exposures. 

On a longer time scale, a large LAr TPC with magnetic
field, e.g. \cite{LAND}, could simultaneously test $\nu_e$ 
appearance and $\nu_\mu$ disappearance for matter effects. 
This and the better energy resolution should in principle
give a good sensitivity. 

\subsection{The LMA solution with high $\Delta m^2_{solar}$} 

Important distortions in the sub-GeV range of the spectrum of
atmospheric electron neutrinos are predicted by the Large Mixing 
Angle (LMA) solution of the solar neutrino problem, provided
$\dm_{solar}$ is high. This scenario, that would also be affected 
by matter effects, has been advocated as a possible solution
of the normalization problem of sub-GeV electrons in Super-Kamiokande
\cite{Peres,Maltoni}. For exactly maximal $\sin^22\theta_{23}$, the
observation of such distortions would be a clean indication of a non
vanishing $\theta_{13}$. 

In large water Cherenkov detectors, the effect could become 
statistically visible. However, the zenith angle dependence is totally
smeared by the experimental resolution at low energies and a
precise knowledge of the absolute normalization of the atmospheric
neutrino flux would be required.
This limitation might be overcome in LAr TPC's, which are
sensitive also to the hadrons produced in the neutrino interactions
and could reconstruct the neutrino direction with a resolution
limited only by Fermi motion and nuclear effects. Still, the effects 
are tiny, and a decisive measurement 
could be made only with very large detectors.

\section{Conclusions and acknowledgements}

Future projects on atmospheric neutrinos could perform precision 
measurements of the oscillation parameters and could provide a direct 
proof of the oscillation mechanism through the explicit observation of
the (as-yet elusive) oscillation pattern in the $L/E$ spectrum of 
atmospheric muon neutrinos. The hierarchy of the neutrino mass 
spectrum could be determined through the observation of matter effects
is a massive (magnetised) detector. The time scale for these projects
is not shorter that ten years.  

I am indebted to G. Battistoni, C.K. Jung, M. Shiozawa, 
R. Gandhi, A. Geiser, P.J. Litchfield, A. Marchionni, S. Ragazzi and 
many others for their help.
I also thank the organizers of the Neutrino 2002 conference for
their hospitality.

\end{document}